\documentstyle[preprint,pre,aps,eqsecnum,epsf]{revtex}

\begin{document}
\draft


\title{Caloric Curves for small systems in the Nuclear Lattice Gas Model}

\author{C. B. Das and S. Das Gupta}
\address{
Physics Department, McGill University,
3600 University St., Montr{\'e}al, Qu{\'e}bec \\ Canada H3A 2T8\\ }

\date{ \today } 

\maketitle

\begin{abstract}
For pedagogical reasons we compute the caloric curve for 11 particles
in a $3^3$ lattice.  Monte-Carlo simulation can be avoided and
exact results are obtained.  There is no back-bending in the caloric
curve and negative specific heat does not appear.  We point out that
the introduction of kinetic energy in the nuclear Lattice Gas Model
modifies the results of the standard Lattice Gas Model in a profound way.
\end{abstract}


\pacs{25.70.Pq, 24.10.Pa, 64.60.My}

\section{INTRODUCTION}
In a recent paper \cite{Das}, we pointed out that microcanonical calculations
in the Lattice Gas Model (LGM) with constant energy are no harder to implement 
than canonical calculations with constant temperature.  We will call the 
first MLGM, and the second, CLGM.  For practical cases at hand $(A\approx 100
$ or 200), the calculations use Monte-Carlo simulations with 
Metropolis algorithm.  We found that in LGM, as used in nuclear
disintegration problems, there is no ``backbending'' in the caloric curve
for systems as small as $^{84}$Kr whether in microcanonical 
or canonical treatments.  By ``backbending'' one means an ``S'' shape
when energy is plotted along the y-axis and temperature along the x-axis.
Since microcanonical treatments seem to lead to backbending for small
systems (100 particles is small enough) in other models 
\cite {Carmona,Gross,Chomaz}, 
our findings need some clarification.  Motivated by this, we present here 
results for a very small system, 11 particles in $3^3$ lattice.  Here we can
avoid Monte-Carlo samplings and do exact (though it still requires some
numerical work which is easy) computations.  The results are quite
interesting and not only explain our previous findings but also shed
light on several connections between microcanonical and canonical
calculations.

\section{The Model System}
As our objective is solely pedagogical, we assume there is just one
kind of particles (nucleons).  We take the number of particles
to be 11.  The
lattice space is $3^3$.  This then implies a freeze-out density 0.41$\rho_0$
which is somewhat higher than the freeze-out density used in lattice
gas model calculations \cite{Pan95}.  The nearest neighbour bonds are
attractive: $\epsilon=-5.33$MeV to get the nuclear matter binding 
energy correct.

The nuclear Lattice Gas Model which is denoted here by LGM is an extension
of the standard textbook Lattice Gas Model as discussed, for example,
in \cite{Huang}.  We denote the standard lattice gas model by SLGM.
The difference is simple: in SLGM, the nucleons are frozen in their
lattice sites.  In LGM, dictated by the physics of the nuclear problem,
they are given momenta.  In CLGM, these momenta are generated using
a Maxwell-Boltzmann distribution.  In MLGM, they are taken from a uniform
distribution within a sphere in momentum space.  The addition
of kinetic energy, however,
changes the caloric curve in an interesting and profound way.  We will
find it useful to discuss the caloric curves in both SLGM and LGM.
Chronologically, it is easier to discuss SLGM first, then point out
how LGM modifies the results.  In both the models the key quantities are
$G(27,11,N_{nn})\equiv g(N_{nn})$= 
the number of configurations with $N_{nn}$ nearest
neighbour bonds for the case of 11 particles in $3^3$ lattice sites.
Once these are known both canonical and microcanonical calculations
are readily done.  The degeneracy factors are given in the small table.
They can be obtained with little effort in this simple case.

\begin{center}
\begin{tabular}{|cc||cc|}
\hline
\multicolumn{1}{|c}{$N_{nn}$} &
\multicolumn{1}{c|}{$g(N_{nn})$} &
\multicolumn{1}{|c}{$N_{nn}$} &
\multicolumn{1}{c|}{$g(N_{nn})$} \\
\hline
0 & 462 & 9 & 2643624 \\
1 & 888 & 10 & 1895907\\
2 & 8511 & 11 & 1051632\\
3 & 38128 & 12 & 481610\\
4 & 150030 & 13 & 174408\\
5 & 481368 & 14 & 50301\\
6 & 1171492 & 15 & 8984\\
7 & 2106504 & 16 & 1056\\
8 & 2772894 & 17 & 96\\
\hline
\end{tabular}
\end{center}
\medskip
\centerline {\it {Table I: Degenracy factors $g(N_{nn})$ with $N_{nn}$ nearest
neighbour bonds.}}

\section{ Microcanonical Treatment of SLGM}
Instead of writing $g(N_{nn})$ we will find it convenient to write
$g$ as a function of $E^*$ where $E^*$ is the excitation energy.
The degeneracy factor $g(E^*)$ as a function of $E^*/|\epsilon|$ is
plotted in Fig. 1.  The distribution is discrete but in Fig.1 we
show it as a continuous distribution and label the y-axis by 
$dN(E^*)/d|\epsilon|$.  If one wants to define a temperature, the
standard practice in the microcanonical model is to compute
$\frac{\partial ln \Omega(E^*)}{\partial (E^*)}\equiv \frac{1}{T}$
(see \cite {Reif}).  An inspection of Fig.1 shows that as a function
of excitation energy the temperature will rise first, approach $+\infty$,
will then switch towards $-\infty$ and as the excitation energy will
further increase the temperature will approach 0 from the negative side.
This happens because in SLGM there is an upper bound to energy.  This
is of course well-known for spin 1/2 systems in a magnetic field if
the kinetic energy of the spin system is suppressed \cite {Reif}.  In
nuclear shell model, for example, this will happen if one restricts 
oneself to limited 
shell model orbitals.  This is well-known to practitioners \cite {Z}.

The caloric curve in microcanonical SLGM is shown in Fig. 2.  In plotting this
curve we used degeneracies of successive discrete points in the 
excitation energy and divided by $|\epsilon|$ to get the temperature.
Notice that in the positive side of the temperature there is no
anomalous behaviour.  If we plot $E^*$ along the y-axis and $T$ along the
x-axis, there is a ``giant'' size backending at about half the excitation
energy available to the system.  But this is merely a reflection of the
fact that the excitation energy available to the system is finite.  This
will drastically change in the nuclear LGM where availability of kinetic
energy will remove the upper limit.

\section{Canonical Treatment of SLGM}
For canonical calculation, we pick a positive temperature: to get the
caloric curve we compute $<E>=\sum \epsilon N_{nn}\times g(N_{nn})\exp(-\beta
N_{nn}\epsilon)/\sum g(N_{nn})\exp (-\beta N_{nn} \epsilon)$.  Subtracting
out the ground state energy we obtain the plot
in Fig. 2.  The same procedure can be used for negative temperature.  Both
are used in Fig. 2.  The similarity between caloric curves calculated
in the microcanonical and canonical models is obvious although there
are quantitative differences.

\section{Microcanonical caloric curve in nuclear LGM} 
From SLGM we now turn to nuclear LGM which serves as a model for
nuclear disaasembly.  This was the case presented in \cite{Das}.
The excitation energy can come from  two sources now: kinetic and potential.
Consequently, we compute $\sum_i g(E_i^*)\rho_{kin}(E^*-E_i^*)$ where 
$g(E_i^*)$ is discrete and taken from the table and $\rho_{kin}(E_{kin})$
is taken to be the integral
\begin{eqnarray}
\int\delta (E_{kin}-\sum p_i^2/2m)\Pi d^3p_i=\frac{(\sqrt{\pi})^{3N}}
{\Gamma(3N/2)}(2m)^{3N/2}E^{3N/2-1}
\end{eqnarray}
$N$ in our chosen case is 11. Now there is no upper limit to
$E^*$.  In Fig. 3 we have plotted $\sum_i g(E_i^*)\rho_{kin}(E^*-E_i^*)$.  The
most important difference from Fig. 1 is that the negative temperature
zone has completely disappeared.  Thus the difference in the caloric curves
obtained from SLGM and LGM will be profound.

There are two ways one can calculate the temperature in the microcanonical
model.  One is the standard 
formula : $\frac{1}{T}=\frac{\partial ln \Omega(E^*)}{\partial E^*}$
where 
\begin{eqnarray}
\Omega(E^*)\propto \sum_i g(E_i^*)\rho_{kin}(E^*-E_i^*)
\end{eqnarray}
The other intuitive approach would be to make the following ansatz.  Although
we are talking of one system only, formally eq. (5.2) is similar to
that of two systems characterised by
$g$ and $\rho_{kin}$ which share energy with each other but are insulated from
the rest of the universe so that the total energy $E^*$ does not change.
If the systems characterised by $g$ and $\rho$ are large
then the sum above would be dominated by the largest term in the sum
which is obtained when the temperature of each subsystem is the same, i.e.,
$\frac{\partial lng(E_i^*)}{\partial E_i^*}=
\frac{\partial ln\rho_{kin}(E_{kin})}{\partial E_{kin}}$.  We now
use $\frac{1}{T}=\frac{\partial ln\rho_{kin}(E_{kin})}{\partial E_{kin}}$.
This leads to
$<T>=<E_{kin}>/(1.5N-1)$.  This $<T>$ and the standard 
definition of $T$  agree quite well as
can be seen in Fig.4.  Notice also there is no backbending in the
microcanonical caloric curve.  If one wants to use the microcanonical
nuclear LGM for practical calculations with nucleon numbers about
100 or higher and also wants to obtain a value for temperature,
getting the temperature from kinetic energy is the only easy option.

In Fig.4 we have also shown the caloric curve in nuclear LGM in the
canonical model.  This agrees with the microcanonical calculation quite
well.

\section{ A saddle-point calculation}
In the particular example (11 particles in $3^3$ boxes in the nuclear
LGM), one has exact expressions for microcanonical density of states.
One can also compute numerically the canonical partition
function.  In nuclear physics one often  has numerical values
for canonical or grand canonical partition functions.
The direct expression for the microcanonical density of state is usually
intractable and in order to obtain a value one uses the saddle-point 
approximation \cite{Bohr,Bhatta}.  We can use the nuclear LGM to
test the accuracy of the saddle-point approximation since here
both the microcanonical density of state and the canonical partition
function are directly calculable.
The microcanonical density of states and the canonical partition
function are related by $Q(\beta)=\int \exp(-\beta E)\rho(E)dE$.
The inverse transformation is $\rho(E)=\exp(\beta_0E)\frac {1}{2\pi}
\int \exp(i\beta E)Q(\beta_0+i\beta)d\beta$.  The saddle-point approximation
for this integral leads to
\begin{eqnarray}
\rho(E)\approx \frac{\exp[\beta_0E+ln Q(\beta_0)]}{\sqrt{2\pi (<E^2>-<E>^2)}}
\end{eqnarray}
where the value of $\beta_0$ is so chosen that at this value $<E>=E$.  The
saddle-point approximation for the density of states is also compared
to the exact density of state in Fig.3.  Except for low excitation energies,
the saddle-point approximation is seen to be excellent.

\section{Summary}
We performed an exact microcanonical calculation of the caloric curve of 
11 particles in a $3^3$ lattice.  The caloric curve does not have a
backbending which means there was no negative specific heat in this
model for 11 particles. We then conclude that the phenomenon of 
backbending can be quite model dependent.

\section{Acknowledgments}
This work was supported in part by the Natural Sciences and Engineering 
Council of Canada and by {\it le Fonds pour la Formation de chercheurs
et l'Aide \`a la Recherche du Qu\'ebec}.  We acknowledge communications
with Professor Dieter Gross.

\newpage

\epsfxsize=4.0in
\epsfysize=6.0in
\par
\centerline{\epsffile{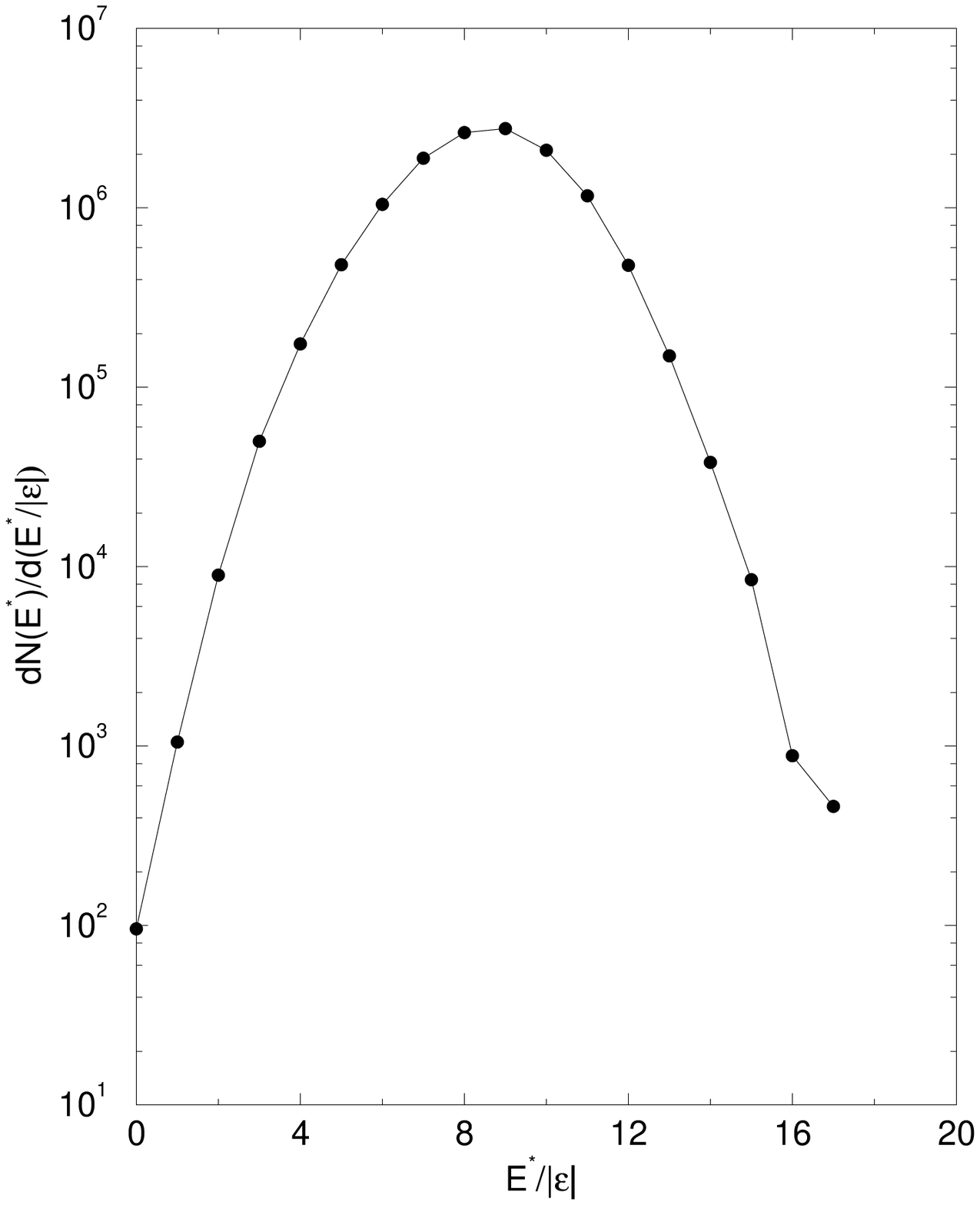}}
\par
\vspace{0.1in}
\it {Fig. 1: The density of states in the standard lattice gas model.
This can be directly obtained from the table remembering that $N_{nn}=17$
defines the ground state.}

\newpage

\epsfxsize=3.5in
\epsfysize=5.0in
\par
\centerline{\epsffile{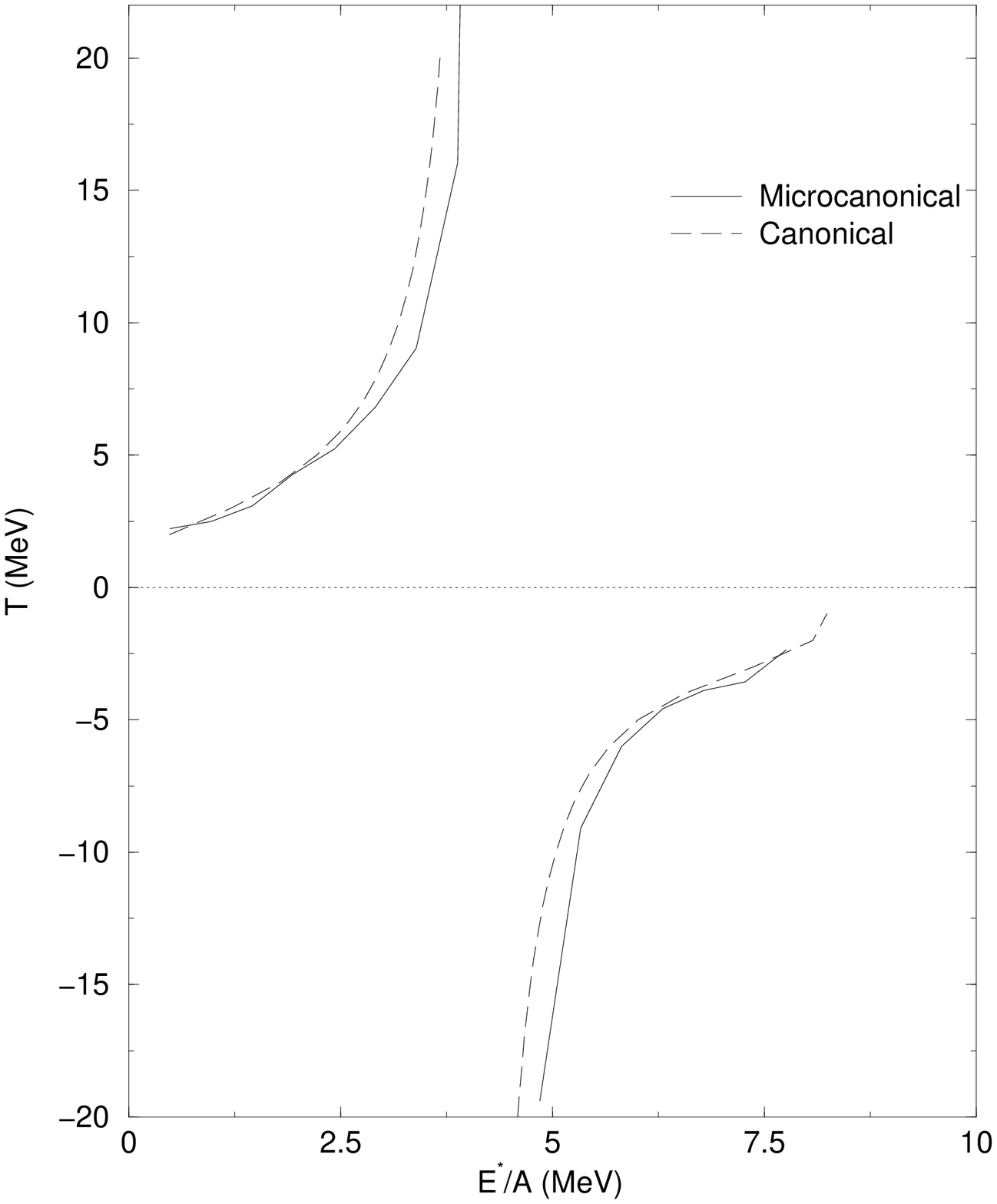}}
\par
\vspace{0.1in}
{\it Fig. 2: The caloric curve in SLGM.  From Fig. 1 it is clear that
the microcanonical defintion of temperature would tend to infinity
around $E^*/|\epsilon| \approx 8$.  For 11 particles this corresponds
to about 4 MeV excitation per particle.  At higher excitations, the
standard definition of temperature leads to large negative temperature.
In the canonical calculation, we assume a temperature (positive
and negative) and obtain $<E^*/A>$ using the table.}

\newpage

\epsfxsize=3.5in
\epsfysize=5.0in
\par
\centerline{\epsffile{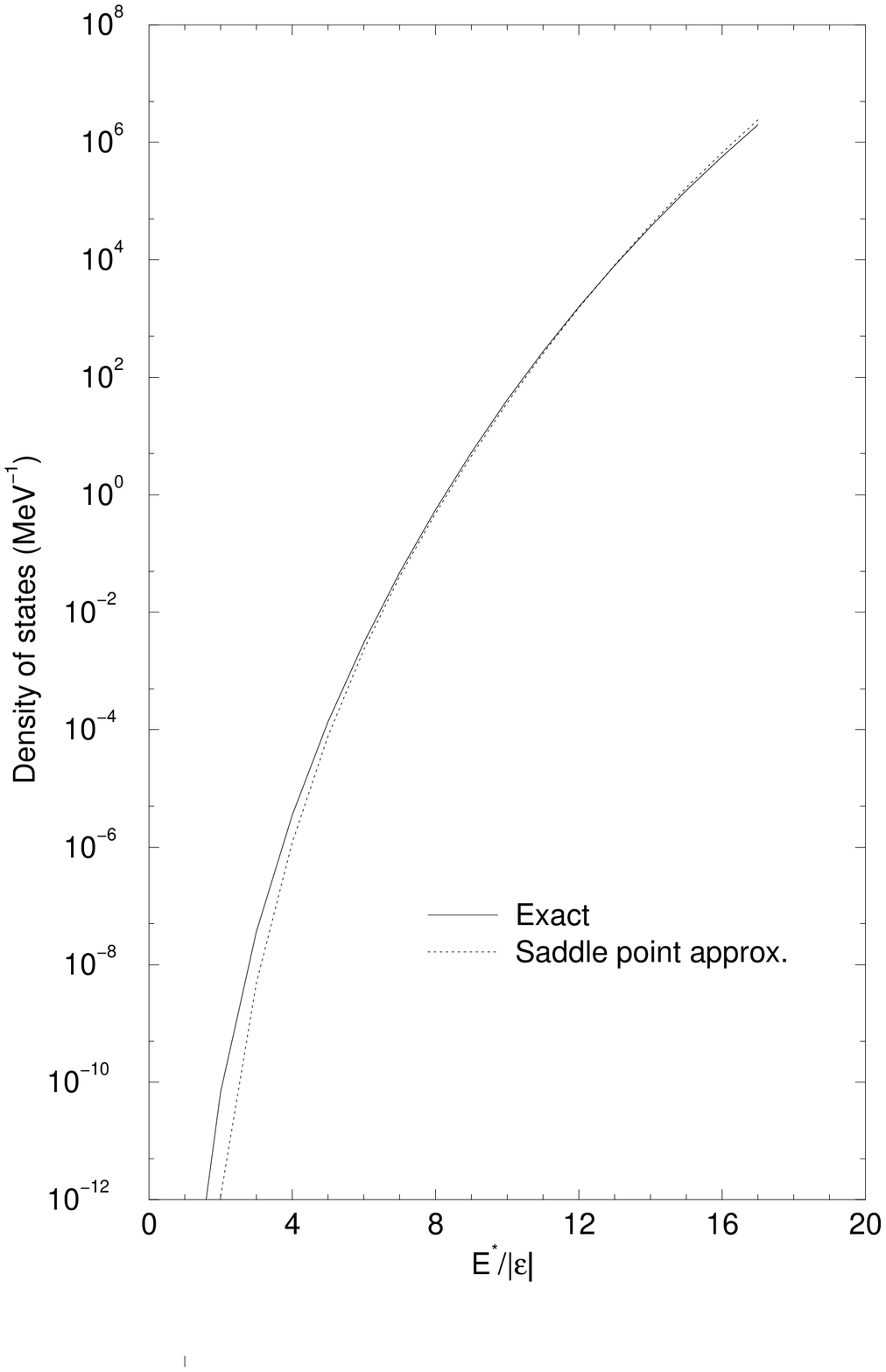}}
\par
\vspace{0.1in}
{\it Fig. 3: The density of states in the nuclear LGM.
We have plotted (the solid curve) $\sum_ig(E_i^*)\rho_{kin}(E^*-E_i^*)$.
For $\rho_{kin}$ we have used eq. (5.1) and multiplied it by
${(\frac{V}{h^3})}^N$ where $V = \frac{27}{0.16} fm^3$.  The dotted curve is the
saddle-point appoximation for the same density of states.  Here $Q(\beta _0)$
is separable into two parts.  One part comes from the potential and is
directly calculable from the table.  This is multiplied by
$(2\pi mT)^{3N/2}$ which comes from the kinetic energy.}

\newpage

\epsfxsize=4.0in
\epsfysize=6.0in
\par
\centerline{\epsffile{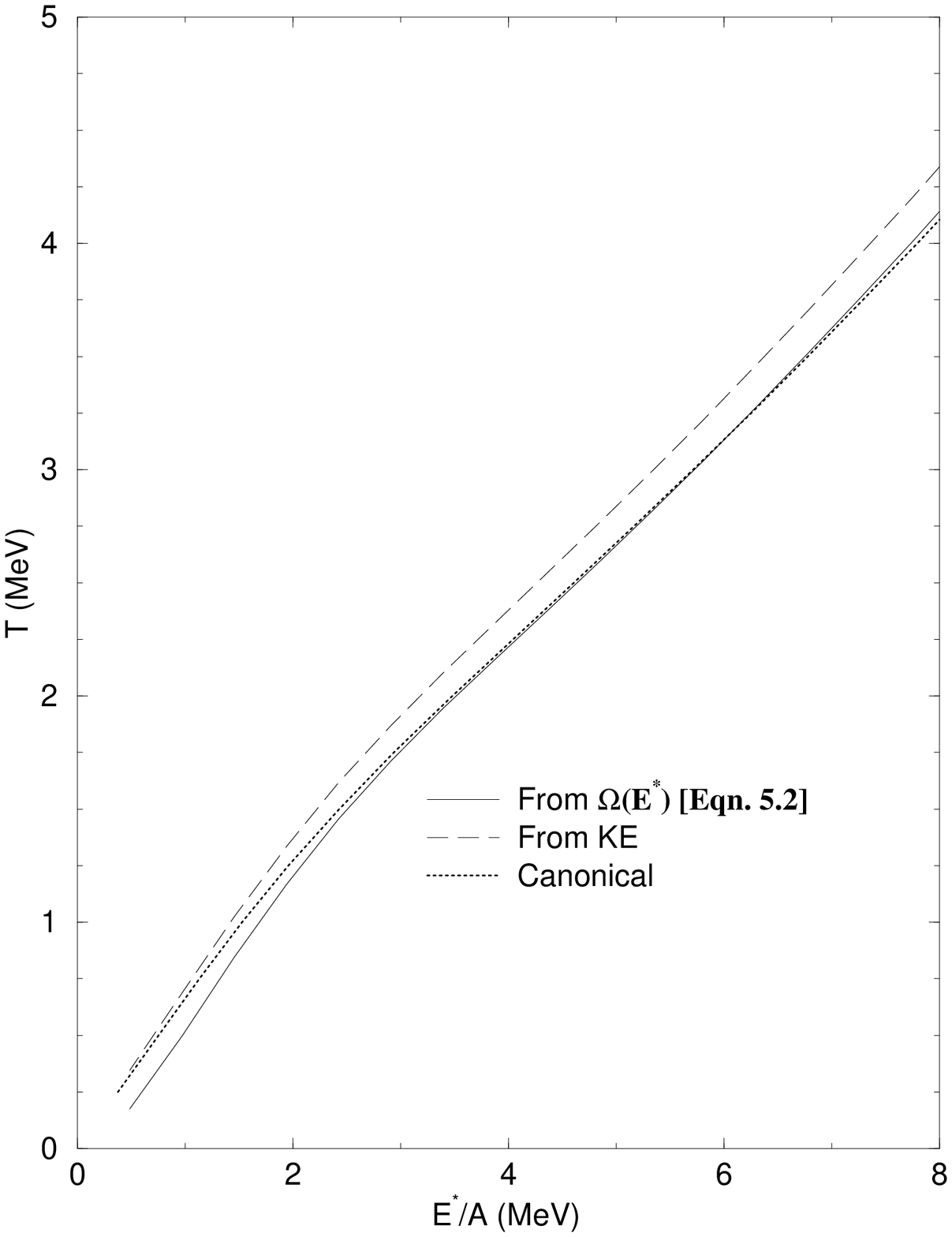}}
\par
\vspace{0.1in}
{\it Fig. 4: The caloric curve in microcanonical and canonical
treatments.  For
microcanonical we show two curves.  The solid curve takes the log of eq. (5.2)
and differentiates with respect to $E^*$ to obtain a temperature.  The
dashed curve defines $T$ from the average value of kinetic energy (see text).
The dotted curve is the canonical caloric curve for the nuclear LGM.}

\end{document}